\documentclass{article}

\usepackage[sglblindworkshop, final]{neurips_2026}
\workshoptitle{Interpreting Agent Behavior (IAB)}

\usepackage[utf8]{inputenc}
\usepackage[T1]{fontenc}
\usepackage{hyperref}
\usepackage{url}
\usepackage{booktabs}
\usepackage{amsfonts}
\usepackage{amsmath}
\usepackage{nicefrac}
\usepackage{microtype}
\usepackage{xcolor}
\usepackage{graphicx}
\usepackage{caption}
\usepackage{tcolorbox}
\usepackage{fontawesome5}
\usepackage{titlesec}
\titlespacing\section{0pt}{6pt plus 2pt minus 2pt}{3pt plus 1pt minus 1pt}
\titlespacing\subsection{0pt}{5pt plus 2pt minus 2pt}{2pt plus 1pt minus 1pt}
\titlespacing\paragraph{0pt}{3pt plus 1pt minus 1pt}{6pt}
\title{Game-Theoretic Solvers with Bayesian Opponent Modeling and Moral Rhetoric Framing for GLEE}

\author{%
  Hayden Arnold \\
  KAIST \\
  \texttt{harnold@kaist.ac.kr} \\
}

\begin{document}

\maketitle

\begin{abstract}
We describe the agent we submitted to the GLEE Competition, which plays bargaining, negotiation, and persuasion games against a shared pool of human and AI opponents. The agent is built around three central components: a classical game-theoretic solver to compute numeric moves, an online Bayesian layer to estimate opponent archetypes, and a language model to write accompanying messages. We utilize Rubinstein bargaining for bargaining, Faratin concession for negotiation, and Kamenica--Gentzkow disclosure for persuasion, and adapt those models to live opponents using two separate layers: a Bayesian estimate of the opponent's hidden discount factor, reservation value, or signaling reliability, and a Thompson sampling bandit over five moral-foundation framings coupled with an evidence-intuition spectrum to establish message register. Most of the development effort went into the recurring problem that theoretically optimal policies perform poorly in certain live settings; we report associated failure modes, guardrails we built, and our hypotheses. We additionally evaluate our model over 13{,}200 Monte Carlo games against 11 synthetic opponent archetypes, an ablation of the opponent-modeling layer, and an extensive analysis of 9{,}337 live-game messages in which moral rhetoric framing in the persuasion setting is strongly correlated with whether the buyer purchased ($\chi^2=616.7$, $p \approx 3.7\times10^{-132}$), surviving controls for quality, round, and configuration.
\end{abstract}
\centerline{\href{https://github.com/haydenarnold/GameTheoreticSolvers}{\faGithub}}

\section{Introduction} \label{sec:introduction}

The GLEE Competition scores agents on the economic outcomes of bargaining, negotiation, and persuasion games against a shared pool of human and AI opponents~\citep{shapira2024glee}. Each game family requires an agent to produce numerically sound decisions and fluent strategic language simultaneously. Our agent handles these requirements through a modular architecture: a classical game-theoretic solver computes numeric decisions, and a language model subsequently generates messages accompanying moves.

The solver for each family is grounded in an established theoretical model. Bargaining follows Rubinstein alternating-offers with online estimation of the opponent's discount factor. Negotiation uses Faratin time-dependent concession, with the opponent's reservation value updated from hard price rejections. Persuasion implements the Kamenica--Gentzkow disclosure rule, modified by an online estimate of buyer trust in the agent's signals. The language module selects a rhetorical framing from an evidence-intuition spectrum and five Moral Foundation Theory arms via Thompson sampling, persisting arm preferences across repeated encounters with the same opponent when identity is disclosed.

While the classical solvers themselves presented few implementation challenges, the dominant challenge was adapting them to opponents whose behavior fell outside the theoretical assumptions of each solver. Several policies that are optimal under their own assumptions produced substantial payoff losses in live games against counterparts who violated those assumptions. We developed guardrails for each such case and validated them on large replays of completed games. 

Overall, we contribute a substantial case study of building and evaluating a game-theoretic agent for economic games, showing how classical equilibrium-based strategies can fail when opponents violate their underlying assumptions. We identify concrete failure modes in bargaining, negotiation, and persuasion, and show how opponent modeling, targeted guardrails, and rhetorical framing can mitigate some of these failures while also introducing new trade-offs.

\section{Related Work} \label{sec:related-work}

GLEE~\citep{shapira2024glee} scores agents on the economic outcome of bargaining, negotiation, and persuasion games. Each family corresponds to a classical model under complete information: alternating-offers bargaining~\citep{rubinstein1982perfect}, a refinement of the axiomatic solution of \citet{nash1950bargaining}; concession-based bilateral trade~\citep{faratin1998negotiation}; and the disclosure problem~\citep{kamenica2011bayesian}. All three assume the opponent's type is known or drawn from a common prior~\citep{harsanyi1967games}. GLEE opponents do not supply this information; we estimate the relevant quantity online from observed behavior. \citet{albrechtstone2018survey} survey opponent-type inference and identify non-stationarity as a core difficulty for approaches that must converge within a single interaction; our shrinkage updates limit the influence of any single observation to address this. \citet{lewis2017dealornodeal} learn to produce offers and messages jointly; we generate them separately, following the staged approach of \citet{hua2024gametheoreticllm}. \citet{bianchi2024negotiationarena} find that LLM negotiation outcomes depend strongly on surface message tactics, suggesting the value of keeping numeric decisions in a separately grounded component. For the rhetoric layer, Moral Foundations Theory~\citep{graham2009liberals} supplies five fixed appeals---care, fairness, loyalty, authority, and purity---making the choice of framing a finite-armed bandit that we solve with Thompson sampling~\citep{chapelleli2011thompson}.

\section{Methodology} \label{sec:methodology}

The agent runs one strategy function per game family behind a dispatcher that routes each game to the right one. Within a family, a classical solver produces numeric moves while a language model produces accompanying messages. Derivations and guardrails for degenerate cases are in Appendix~\ref{sec:appendix-methodology}.

\subsection{The decision layer}
\label{sec:decision}

\paragraph{Bargaining.} Two players alternate proposing how to split a fixed sum, and delay is costly because per-round interest may be collected. Following \citet{rubinstein1982perfect}, under complete information and an unlimited horizon the stationary subgame-perfect equilibrium gives the proposer a share $x^\ast = (1-\delta_{\mathrm{opp}})/(1-\delta_{\mathrm{me}}\,\delta_{\mathrm{opp}})$ of the pot, where $\delta_{\mathrm{me}}$ and $\delta_{\mathrm{opp}}$ are the per-round discount factors of the proposer and responder, each a number in $(0,1)$ saying how much a payoff is worth one round later (a factor of $0.9$ means ten percent is lost per round). When $\delta_{\mathrm{opp}}$ is near one, the opponent has weak incentives to concede early, and the proposer's equilibrium share falls. A finite, known horizon uses backward induction from the deadline instead.

\paragraph{Negotiation.} A seller and buyer alternate price offers for one product, each holding a private value, with no per-round decay; the only pressure is the round cap and the option to walk away. We set the asked price by the time-dependent tactic of \citet{faratin1998negotiation}, $\mathrm{price}(t) = r + (1 - t^{1/\beta})(a - r)$, where $r$ is our reservation value (a seller's cost or a buyer's maximum), $a$ is our opening anchor, $t \in [0,1]$ measures how far the negotiation has run (the fraction of the round cap used, or $1 - e^{-\mathrm{round}/\tau}$ when unlimited, with $\tau$ setting the pace), and $\beta$ shapes the concession: below one holds firm early and concedes near the deadline (Boulware), above one concedes early then holds. We update the opponent's value interval only on hard price rejections; a rejected price establishes a verified bound on the opponent's value. Counteroffers do not carry equivalent information and are excluded from this update.

\paragraph{Persuasion.} A seller offers a product each round at a fixed price, worth $v$ to the buyer with prior probability $p$ (high quality) and $u < \mathrm{price}$ otherwise; the seller sees the true quality, the buyer sees only $p$ and the history. Following \citet{kamenica2011bayesian}, the seller signals ``buy'' on high quality and, on low quality, bluffs with probability $q^\ast = p\,(v - \mathrm{price}) / [(1-p)\,(\mathrm{price} - u)]$, clipped to $[0,1]$. This is the bluff rate at which a buyer who receives a ``buy'' signal is indifferent between buying and passing, given the buyer's knowledge of how often the seller bluffs. Exceeding $q^\ast$ renders the signal uninformative to a rational buyer; falling below it forgoes achievable sales. The formula assumes a memoryless buyer; we revisit this assumption in Section~\ref{sec:behavior-analysis}.

\subsection{Adapting to an unknown opponent} \label{sec:opponent-modeling}

Each solver requires a quantity the opponent does not reveal. In bargaining we maintain a distribution over a grid of candidate discount factors and update it each round the opponent proposes. We compare the opponent's self-ask to the self-ask predicted by each candidate and reweight by $w_i \leftarrow w_i \cdot \exp[-(\hat{x}(\delta_i) - x_{\mathrm{obs}})^2 / (2\sigma^2)]$, where $w_i$ is the current weight on candidate factor $\delta_i$, $\hat{x}(\delta_i)$ is the self-ask that candidate predicts for the current round, $x_{\mathrm{obs}}$ is the self-ask the opponent made, and $\sigma$ sets how sharply a mismatch is penalized. The posterior mean gives $\delta_{\mathrm{opp}}$ for the solver. After each update we shrink the new weights back toward their pre-update values by a factor that decreases as updates accumulate, which limits the influence of any single offer on the estimate; when the opponent's self-ask has stopped decreasing over several consecutive offers, we apply that observation a second time so that the shrinkage catches up with the confidence the repetition has already earned (Appendix~\ref{sec:appendix-methodology}). In negotiation the opponent's value is bounded by a running interval that tightens on hard price rejections. As the persuasion buyer, the agent estimates the probability of high quality given its own purchase history by $\hat{p} = (k\,p + h)/(k + n)$, where $n$ is the number of purchases, $h$ is the number that were high quality, $p$ is the prior the game supplies, and $k$ is a pseudo-count that controls how strongly the prior is weighted before observed purchases dominate.

\subsection{The message layer} \label{sec:message-layer}

We use a language model to generate an accompanying message for each move, up to two thousand characters. We model opponent profiles on two components: whether the opponent argues using evidence-based or intuition-based language, and which of the five moral foundations they lean on. We then condition the framing of replies on this profile. A Thompson sampling bandit over those five foundations selects the framing to lead with: each arm maintains a Beta posterior over its success rate, the agent draws from each posterior and plays the highest draw, then updates the chosen arm on the observed outcome. When opponent identity is disclosed, which occurs in roughly half of games, the bandit state is stored per opponent and reloaded on the next encounter. We describe the message layer methodology in more detail in Appendix~\ref{sec:appendix-rhetoric}, with qualitative examples in Appendix~\ref{sec:appendix-counter-framing}.

\section{Evaluation} \label{sec:evaluation}
We evaluate the agent on the live leaderboard, in a Monte Carlo suite against synthetic opponents, in an ablation of the opponent-modeling layer, and in an analysis of rhetorical framing in live games.

\begin{figure}[t]
  \centering
  \includegraphics[width=1\linewidth]{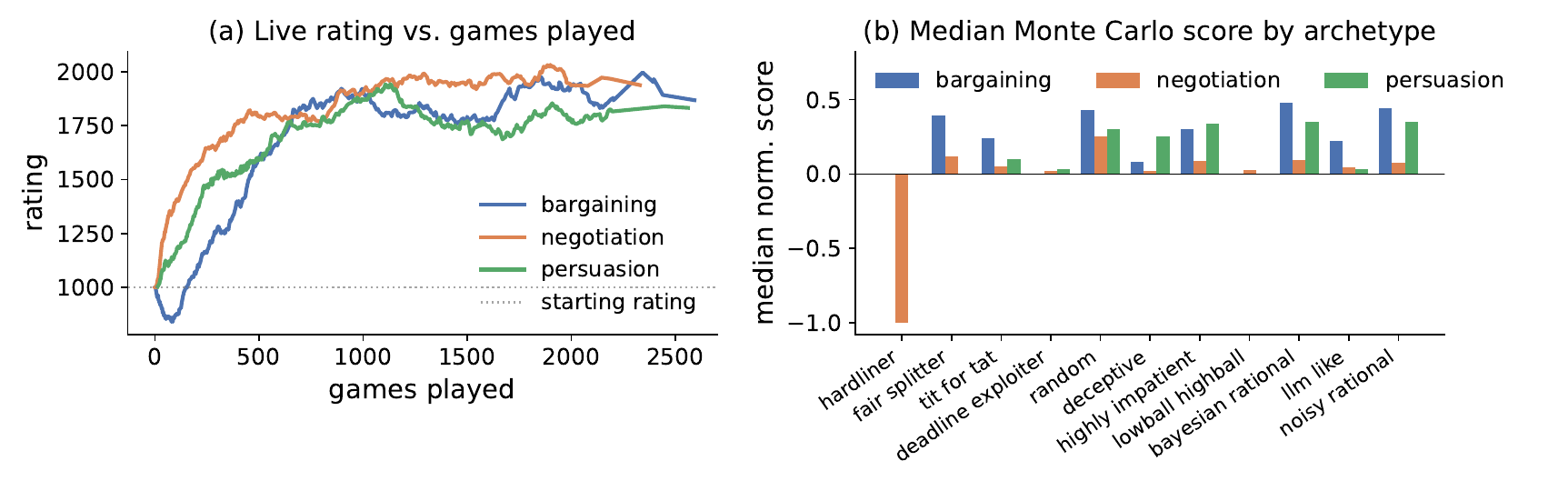}
  \caption{(a) Displayed rating against the first 2600 live games played, per family; the dotted line is the $1000$ start. (b) Median normalized Monte Carlo score per archetype. The one clearly negative bar, \texttt{hardliner} in negotiation ($-1.0$), holds a fixed position throughout the game that is strongly unfavorable to both players.}
  \label{fig:rating}
\end{figure}

\paragraph{Live standing.} GLEE maps each game's opponent-adjusted payoff onto a rating scale centered at $2000$, the field median, with the displayed rating pulled toward the $1000$ starting value by a factor $g/(g+30)$ where $g$ is games played in that family; displayed ratings converge to underlying ratings as games accumulate. The agent reached peak displayed ratings of $1996.4$ in bargaining (at $2{,}343$ games), $2131.1$ in negotiation (at $5{,}418$), and $1941.9$ in persuasion (at $1{,}130$), and finished $55$th of $198$ teams on the aggregate leaderboard that averages the three families (Figure~\ref{fig:rating}a).

\paragraph{Synthetic opponents.} We built $11$ archetypes spanning a range from the \texttt{hardliner}, which holds a fixed extreme position, to the \texttt{bayesian\_rational}, which best-responds to public information, and ran the deployed code against each for $400$ trials per family ($13{,}200$ games; Appendix~\ref{sec:appendix-archetypes}). Median score falls as an archetype's behavior departs from the assumptions the relevant solver targets (Figure~\ref{fig:rating}b; full tables in Appendix~\ref{sec:appendix-tables}): the \texttt{bayesian\_rational} yields our highest medians, for example $+0.478$ in bargaining, while the \texttt{hardliner} in negotiation is the only negative median ($-1.0$), since agreement there requires our reservation value to fall inside a fixed demand, which randomized value assignment rarely produces. Negotiation means sit well below their medians because roughly half of trials leave no zone of agreement and floor at $-1$.

\paragraph{Ablation.} Disabling the three opponent-estimation modules and replacing them with fixed priors on identical seeds leaves bargaining and negotiation aggregates inside their per-cell noise ($+0.238$ to $+0.234$ and $-0.251$ to $-0.262$) (Tables~\ref{tab:ablation_main}, \ref{tab:ablation}). Persuasion is the exception: removing the buyer-trust model raises the mean from $+0.269$ to $+0.312$, so the module we deployed costs average-case payoff in this suite. Section~\ref{sec:behavior-analysis} interprets this alongside deceptive-seller behavior we sought to inhibit.

\begin{table}[h]
\centering
\small
\caption{Mean normalized Monte Carlo score over the 11 archetypes, with the opponent-estimation modules enabled (Full) and replaced by fixed priors (Ablated), on identical seeds (400 trials per archetype).}
\vspace{5pt}
\label{tab:ablation_main}
\begin{tabular}{lccc}
\toprule
Family & Full & Ablated & $\Delta$ (Ablated $-$ Full) \\
\midrule
Bargaining  & $+0.238$ & $+0.234$ & $-0.004$ \\
Negotiation & $-0.251$ & $-0.262$ & $-0.011$ \\
Persuasion  & $+0.269$ & $+0.312$ & $+0.043$ \\
\bottomrule
\end{tabular}
\end{table}

\paragraph{Rhetoric framing.} We classified 9,337 outgoing persuasion messages, 4,661 bargaining and 7,558 negotiation messages by dominant moral foundation, and tested independence between framing and outcome per family (Appendix~\ref{app:foundation}). The association is strongly game-dependent. In persuasion, buy rate rises from 30.9\% under care to 65.8\% under loyalty ($\chi^2=616.7$, $p\approx3.7\times10^{-132}$; $\chi^2(4)=263.5$ with standard errors clustered by game). Part of care's deficit comes from the success definition: 50.1\% of care messages are honest ``pass'' messages, which a buyer should decline (Table~\ref{tab:pitch}). Restricted to pitches, the care--loyalty gap narrows from 34.8 to 18.6 points (odds ratio $2.18$ for loyalty against care, controlling for quality and round), and on true bluffs, pitches on low-quality rounds, care reaches 59.6\% against loyalty's 73.5\%. Bargaining shows the same frame ranking over a much narrower spread ($\chi^2=11.2$, $p=0.0249$, Cram\'er's $V=0.05$), and negotiation shows no significant association ($\chi^2=9.1$, $p=0.059$), with acceptance between 2.8\% and 5.1\% regardless of frame. The pattern tracks how directly each family ties its outcome to language: persuasion's outcome is a trust decision the buyer can reach only through the seller's words, negotiation turns on a price already on the table, and bargaining sits between them. The evidence--intuition register shows the same ordering. In persuasion, evidence-leaning messages succeed 62.4\% of the time against 47.6\% for balanced and 44.1\% for intuition-leaning ones (odds ratio $1.83$ per unit of evidence-leaning with controls, $2.11$ on true bluffs), while bargaining ($0.98$) and negotiation ($0.88$) show no association. All of these are correlational, since the bandit chose framings adaptively. Appendix~\ref{sec:appendix-counter-framing} gives qualitative examples in which the message layer adopts an opponent's rhetorical register and reverses its evaluative direction, which we label counter-framing.

\begin{table}[h]
\centering
\small
\caption{Persuasion buy rate by dominant foundation of the seller's message. Pass share is the fraction of the foundation's messages that were honest ``pass'' messages; pitches are the rest, and true bluffs are pitches on low-quality rounds. Intervals are in Appendix~\ref{app:foundation}.}
\vspace{5pt}
\label{tab:pitch}
\begin{tabular}{lrrrrr}
\toprule
Foundation & $n$ & Pass share & All & Pitches & True bluffs \\
\midrule
Loyalty   & 2,667 &  6.3\% & 65.8\% & 68.4\% & 73.5\% \\
Purity    & 2,119 & 11.4\% & 57.3\% & 62.5\% & 68.8\% \\
Fairness  & 2,115 & 16.9\% & 56.1\% & 63.9\% & 74.1\% \\
Authority &   311 &  7.4\% & 49.2\% & 52.1\% & 70.5\% \\
Care      & 2,125 & 50.1\% & 30.9\% & 49.8\% & 59.6\% \\
\bottomrule
\end{tabular}
\end{table}

\section{Agent Behavior Analysis} \label{sec:behavior-analysis}
The competition scores an agent on an opponent-adjusted percentile that stabilizes only over many games, so we validated every change on large replays of completed games rather than on one or two live results. We discovered the same pattern during this process: a policy that is optimal under its solver's assumptions frequently loses to certain opponent behaviors the design had not anticipated. We describe the divergences that drove the most change in our model, the guardrails they produced, and how we keep deployed behavior aligned with design. Figures are in GLEE payoff units; constants are included in Appendix~\ref{sec:appendix-methodology}.

\paragraph{Opponent honesty persists across rounds.} Two components of our persuasion model initially treated each round in isolation, which is correct for product quality but is suboptimal against dishonest opponents. As a seller, the Kamenica--Gentzkow bluff rate $q^\ast$ is derived for a buyer who evaluates each round independently; against live buyers who tracked our signal history, a single discovered bluff suppressed purchasing for the rest of a game at a cost exceeding the round it won. As a buyer, the initial trust model re-explored a seller it had locked out, on the assumption that an unlucky early sample should not permanently condemn a seller, given the independence of product quality from seller behavior. However, over a sample of $102$ live games, such re-exploration purchases landed a high-quality product $4.2\%$ of the time, against $76.3\%$ for purchases from trusted sellers. As a result, we removed re-exploration and capped our own bluff rate in the opening rounds, having tested non-bluffing and more-selective sellers and found that neither improved aggregate replay outcome. This conservative strategy is correctly considered a cost during synthetic testing (Section~\ref{sec:evaluation}), but effectively shielded our agent from deceptive opponents.

\paragraph{A single offer is weak evidence of a type.} In bargaining, one self-ask from the opponent moved our posterior mean over their discount factor from the grid average of $0.819$ to $0.92$, because that observation lay far from any patient candidate, and the jump turned a fair opening into a large concession. The opponents that triggered it were erratic or trend-reversing, offering little stationary signal about a fixed type. We now shrink each posterior update toward its pre-update value by a factor that decays as observations accumulate, which cut mean absolute error against an off-grid rational opponent from $0.053$ to $0.023$ after three consistent observations. The first update is identical to the earlier fixed-prior shrinkage, so the original incident survives as a reproducible unit test.

\paragraph{Discovered archetypes and alignment.} Several of the synthetic archetypes in Section~\ref{sec:evaluation} began as behaviors we first met in live play, among them the history-tracking buyer behind the bluff cap and the trend-reversing bargainer behind the shrinkage step. Reproducing them synthetically confirmed that the divergences above were recurring archetypes rather than isolated games, and it grounds the Monte Carlo gradient: performance falls smoothly as an opponent departs from a solver's assumptions because those departures are the cases the guardrails target. We continued this loop throughout the duration of the competition by replaying recently completed games to locate a divergence, encoding the opponent behavior as both a synthetic archetype and a unit test, and retaining a proposed fix when it improves aggregate replay outcome (Table~\ref{tab:guardrails}). Modifications that did not improve aggregate replay outcome were removed and are recorded in Appendix~\ref{sec:appendix-methodology}.

\begin{table}[h]
\centering
\small
\caption{Divergences between theoretical solver assumptions and live opponents, and resulting guardrails.}
\label{tab:guardrails}
\begin{tabular}{p{3.0cm}p{4.6cm}p{5.1cm}}
\toprule
Divergence & Evidence & Guardrail \\
\midrule
Honesty persists across rounds (seller) & A discovered bluff suppressed purchasing for the rest of a game & Bluff rate capped in the first three rounds; set to zero after three consecutive refusals \\
Honesty persists across rounds (buyer) & Re-exploration purchases were high quality 4.2\% of the time vs.\ 76.3\% for trusted sellers (102 games) & Re-exploration removed \\
A single offer is weak evidence of a type & One self-ask moved the discount-factor mean from 0.819 to 0.92 & Shrinkage toward pre-update weights; MAE against an off-grid opponent fell from 0.053 to 0.023 \\
Raw Rubinstein share degenerates & Near-one discount factors give a whole-pot demand or zero & Degenerate limit routed through the infinite-horizon formula; shares clamped to 30--70\%; finite value damped \\
Holdouts run into a hidden cap & Unlimited games end near round 99--100 & Deadline safety zone: mirror the opponent, accept any net-positive offer at the last chance \\
Stuck negotiations end in walk-aways & No per-round decay, so no pressure to settle & Accept a profitable offer that has stopped moving \\
\bottomrule
\end{tabular}
\end{table}

\section{Limitations}
The framing analysis is correlational: the bandit chose framings adaptively, the language model sets the surface wording, and foundation and register labels are lexical. Standard errors are clustered by game because opponent identity is not recorded in the game index, so repeated opponents are not accounted for. A randomized intervention that fixes the framing would be needed to establish a causal effect. The synthetic archetypes are our own construction, and the ablation is only run against them; live contribution of each opponent-estimation module is not measured. Because identity is disclosed in roughly half of games and opponents are typically met only once or twice, cross-game learning rarely engages. Finally, our live standing was modest: peak ratings were 1996.4 in bargaining and 1941.9 in persuasion, 2131.1 in negotiation, and we finished 55th of 198 teams.

\clearpage

\bibliographystyle{plainnat}
\bibliography{references}

\appendix

\section{Solver Derivations and Guardrails} \label{sec:appendix-methodology}

\paragraph{Bargaining.} The infinite-horizon share $x^\ast$ discussed in Section~\ref{sec:decision} is the stationary subgame-perfect equilibrium of \citet{rubinstein1982perfect}, generalized to asymmetric discount factors $\delta_{\mathrm{me}}$ and $\delta_{\mathrm{opp}}$; the equilibrium share depends on these factors because the more patient player has greater bargaining power. For a finite, known horizon we compute the proposer's take by backward induction from the deadline, where the player who would propose at the final round can claim the whole pot if reached. Pure backward induction oscillates by round parity when a discount factor is near one. We therefore damp the finite value toward the infinite-horizon value using a weight that decays as the deadline recedes, so the exact finite-horizon result dominates only near the deadline. A discount factor within one part in a thousand of one is treated as the degenerate limit and routed through the stable infinite-horizon formula in both the finite and infinite cases. We also cap the agent's proposed and accepted shares between 30\% and 70\% in degenerate or unlimited-horizon games, where the raw formula can produce extreme values. In ordinary finite games, we use the exact share derived by backward induction. A deadline safety zone, entered a fixed number of rounds before a known deadline and near the unlimited-horizon hidden round cap we observed around round $99$ to $100$, has the agent mirror the opponent's last stated position and accept any net-positive offer at the true last chance, which prevents a theoretically justified holdout from ending in the zero-payoff outcome that a missed deadline forces.

\paragraph{Negotiation.} The concession target discussed in Section~\ref{sec:decision} is the time-dependent tactic of \citet{faratin1998negotiation}. The progress variable uses elapsed rounds rather than the current round index, so a one-shot ultimatum does not register as fully time-elapsed on its only move. The opponent-value interval tightens only from a hard rejection, as described in Section~\ref{sec:opponent-modeling}; a counteroffer is treated as a biased signal and does not tighten the bound. When the opponent's offer has been stationary for several rounds and is already profitable, the agent accepts it rather than continuing to push, since this family has no per-round decay and the only way a stuck negotiation ends is a walk-away.

\paragraph{Persuasion.} The bluff rate discussed in Section~\ref{sec:decision} is the disclosure policy of \citet{kamenica2011bayesian} under the game's payoff structure. When the seller does not know the buyer's value $v$, the agent cannot compute $q^\ast$ exactly and instead adjusts its bluff rate toward whatever keeps the observed buyer accept-rate consistent with a credible seller. The trust-building cap of Section~\ref{sec:behavior-analysis} limits the bluff rate in the first three rounds regardless of $q^\ast$.

\paragraph{Belief update.} Each update blends the likelihood-reweighted weights with the pre-update weights according to $\mathbf{w} \leftarrow (1-m)\mathbf{w}_{\mathrm{raw}} + m\mathbf{w}_{\mathrm{prev}}$, where $m = \frac{k}{k+n}$, $k=3$, and $n$ is the number of updates applied so far. Consequently, the first update is pulled $75\%$ toward the uniform prior, and this influence weakens as evidence accumulates. Because $n$ counts updates rather than offers, a single surprising offer is damped even when its likelihood is sharp. If the opponent's last three self-asks have not decreased by more than $0.5\%$ of the pot, and the game involves incomplete information, the current observation is applied a second time within the same turn. This is done because the damping schedule depends on the number of rounds observed, rather than on how consistent those observations have been. Without this second application, belief convergence lags behind an opponent who is already visibly holding out.

\paragraph{Guardrail on seller re-exploration.} We initially had the buyer's trust model re-explore untrustworthy sellers, on the theory that quality is drawn independently each round, so an early unlucky sample should not cause a permanent lockout. Replayed against $102$ completed games, those re-exploration purchases landed a high-quality product $4.2\%$ of the time, against $76.3\%$ for purchases from trusted sellers. The premise was wrong: product quality is independent across rounds, but a seller's honesty is a fixed trait of that opponent, so re-testing a seller already shown to be unreliable simply loses again. We removed the mechanism, and results presented in the paper come from after this removal.

\paragraph{Constants.} Table~\ref{tab:constants} lists the constants of the deployed agent; payoff-valued thresholds are fractions of the pot (bargaining) or of our reservation value (negotiation).

\vspace{-5pt}
\begin{table}[h]
\centering
\small
\caption{Constants for the deployed agent.}
\label{tab:constants}
\scriptsize
\begin{tabular}{p{2.5cm}p{10.6cm}}
\toprule
Component & Constants \\
\midrule
Bargaining belief & Discount grid $\{0.5,0.6,0.7,0.8,0.85,0.9,0.93,0.95,0.97,0.99\}$; kernel $\sigma=0.08$; shrinkage pseudo-count $k=3$. \\
Bargaining moves & Proposed share reduced by 3\%; shares clamped to 30--70\%; accept tolerance 1\% of the pot; open at 50/50 in round 1 when $\delta_{\mathrm{me}}\le0.90$ (incomplete information, unlimited horizon); degenerate limit $\delta\ge0.999$, blended in from $0.99$; concession $\beta=0.7$, $\tau=30$ ($\tau=10$ once a holdout is confirmed); ask lowered by only 30\% of the target drop against a holdout. \\
Bargaining deadline & Safety zone at $\le2$ rounds remaining (finite) or round $\ge90$ (unlimited), with last-chance acceptance of any positive offer at round $97$; inside the zone accept offers of at least 10\% of the pot. Holdout confirmed after 14 non-conceding offers (7 for an opponent previously confirmed hopeless in at least two games). \\
Stationary detection & Three consecutive offers within 0.5\% of the pot (seven when $\delta_{\mathrm{me}}\ge0.999$). \\
Negotiation & $\beta=0.85$; $\tau=30$ when unlimited; prior value spread $1.5\times$; one-shot offer at $t=0.75$ (incomplete information) or $0.03$ (complete); accept at once any offer with at least 30\% margin; accept a stalled profitable offer (three offers within 0.5\% of value, and under complete information at least 10\% of the agreement zone); concession damped by 30\% against a stalled opponent. \\
Persuasion buyer & Safety margin 20\% over price; trust prior weight $k=20$, prior used until two purchases; decline-model prior $\Pr(\text{high})=0.15$; first purchase when prior EV/price $\ge0.95$; test a pass signal when upside/downside $\ge1.2$ and prior EV/price $\ge1.2$; language-model discourage verdict used at confidence $\ge0.7$, otherwise a phrase list; a message repeated for three rounds is treated as uninformative. \\
Persuasion seller & Bluff rate $0.7\,q^\ast$ ($0.35$ base when $v$ is unknown); zero in rounds 1--3; adaptive step $0.05$ over a five-round window (raise if buy rate $>0.8$, lower if $<0.4$); zero after three consecutive refusals. \\
Message layer & Language model: Gemini 3.7 Flash; target about 300 characters (cap 2,000); bandit prior strength 1 per game and 4 when opponent identity is known. \\
Evaluation & Fixed Monte Carlo master seed, 400 trials per archetype per family. \\
\bottomrule
\end{tabular}
\end{table}

\section{Synthetic Opponent Archetypes} \label{sec:appendix-archetypes}

The taxonomy is our own construction, drawing on behavioral types described in the negotiation literature~\citep{raiffa1982art, faratin1998negotiation} and on tit-for-tat as introduced by~\citet{axelrod1984evolution}. Each archetype observes the agent's actual move each round and computes its own response. Bargaining and negotiation archetypes expose a rule for the price or share they propose and a rule for whether they accept; persuasion archetypes expose a rule for the seller's signal or the buyer's purchase.

\begin{itemize}
  \item \textbf{hardliner}: a fixed extreme demand that never concedes, and
  accepts only a large margin in its own favor. As a persuasion seller it always
  signals ``buy''; as a buyer it never purchases.
  \item \textbf{fair\_splitter}: targets an even split or a small fixed margin
  over its value, and accepts any offer at or above that target. Signals honestly
  as a seller; buys when signal and prior jointly clear the price.
  \item \textbf{tit\_for\_tat}: mirrors the agent's most recent concession and
  tightens its accept threshold in proportion to how much the agent has conceded.
  As a buyer it tracks the seller's signal accuracy and buys once trust passes
  one half.
  \item \textbf{deadline\_exploiter}: holds an extreme position until the final
  rounds, then concedes sharply. As a seller it signals ``buy'' unconditionally
  until late in the game, then reverts to honesty.
  \item \textbf{random}: draws its ask, price, and accept threshold uniformly at
  random each round.
  \item \textbf{deceptive}: follows a Boulware curve but reverses upward with
  probability $0.35$ each round, breaking any trend an opponent model would infer.
  Bluffs on low quality with probability one half.
  \item \textbf{highly\_impatient}: a fast conceder that lowers its demand quickly
  and loosens its accept threshold as the game runs. As a buyer it purchases on
  any positive signal.
  \item \textbf{lowball\_highball}: opens at an extreme anchor and concedes very
  slowly, demanding a large margin to accept.
  \item \textbf{bayesian\_rational}: plays the theoretically optimal ask, accept
  threshold, or bluff rate under an assumed symmetric patience parameter; as a
  buyer it buys when a Bayesian-updated expected value clears the price.
  \item \textbf{noisy\_rational}: \texttt{bayesian\_rational} with added Gaussian
  noise on its ask, threshold, or bluff rate.
  \item \textbf{llm\_like}: a moderate conceder that rounds its offers and adds
  small noise, approximating a stylized LLM counterpart. Bluffs with probability
  about $0.3$ regardless of quality.
\end{itemize}

\section{Full Monte Carlo and Ablation Tables}
\label{sec:appendix-tables}

\vspace{-15pt}
\begin{table}[h]
  \caption{Mean (standard deviation) and median normalized score, $400$ Monte
  Carlo trials per archetype per family, on the deployed strategy code with all
  three opponent-modeling mechanisms enabled. For bargaining and persuasion,
  $0$ is the even-split or no-sale baseline; for negotiation, $-1$ is a
  walk-away or no-deal, and the mean sits below the median because roughly half
  of trials have no zone of agreement and floor at $-1$.}
  \vspace{5pt}
  \label{tab:montecarlo}
  \centering
  \scriptsize
  \begin{tabular}{lrrrrrr}
    \toprule
    & \multicolumn{2}{c}{Bargaining} & \multicolumn{2}{c}{Negotiation} & \multicolumn{2}{c}{Persuasion} \\
    Archetype & mean (sd) & median & mean (sd) & median & mean (sd) & median \\
    \midrule
    hardliner           & +0.011 (0.017) & +0.002 & $-0.656$ (0.545) & $-1.000$ & +0.080 (0.232) & +0.000 \\
    fair\_splitter      & +0.385 (0.124) & +0.391 & $-0.192$ (0.636) & +0.121 & +0.288 (0.399) & +0.000 \\
    tit\_for\_tat       & +0.250 (0.183) & +0.240 & $-0.273$ (0.578) & +0.050 & +0.317 (0.387) & +0.100 \\
    deadline\_exploiter & +0.025 (0.051) & +0.003 & $-0.256$ (0.555) & +0.019 & +0.149 (0.244) & +0.034 \\
    random              & +0.429 (0.177) & +0.428 & +0.029 (0.553) & +0.253 & +0.262 (0.230) & +0.300 \\
    deceptive           & +0.122 (0.127) & +0.082 & $-0.329$ (0.589) & +0.024 & +0.335 (0.341) & +0.250 \\
    highly\_impatient   & +0.273 (0.125) & +0.300 & $-0.193$ (0.585) & +0.087 & +0.357 (0.342) & +0.339 \\
    lowball\_highball   & +0.024 (0.046) & +0.003 & $-0.309$ (0.563) & +0.027 & +0.166 (0.318) & +0.000 \\
    bayesian\_rational  & +0.467 (0.165) & +0.478 & $-0.187$ (0.567) & +0.096 & +0.380 (0.358) & +0.350 \\
    llm\_like           & +0.185 (0.131) & +0.224 & $-0.266$ (0.598) & +0.047 & +0.279 (0.372) & +0.033 \\
    noisy\_rational     & +0.447 (0.183) & +0.442 & $-0.131$ (0.540) & +0.073 & +0.344 (0.313) & +0.350 \\
    \bottomrule
  \end{tabular}
\end{table}

The ablation in Section~\ref{sec:evaluation} disables the three opponent-estimation modules—the bargaining discount-factor belief, the negotiation value interval, and the persuasion buyer-trust model—by replacing their updates with fixed baseline priors on identical seeded trials. Bargaining aggregate scores were essentially unchanged across all 11 archetypes ($+0.238$ baseline versus $+0.234$ ablated, a $-0.004$ difference well inside the per-cell noise floor). Negotiation produces visibly different per-archetype behavior (e.g. tit\_for\_tat degrades from $-0.273$ to $-0.322$ once the estimator is disabled), but the aggregate effect remains small: $-0.251$ baseline versus $-0.262$ ablated, a $-0.011$ difference against a per-cell standard deviation of $0.54$–$0.64$ at $n=400$. Persuasion remains the only family where opponent modeling measurably altered performance. Table~\ref{tab:ablation} gives the full per-archetype Full-versus-Ablated breakdown for all three families.

\begin{table}[h]
  \caption{Mean normalized score, full system against opponent-modeling
  ablated, $400$ trials per archetype per family, on identical seeds. The Full columns match Table~\ref{tab:montecarlo}.}
  \vspace{5pt}
  \label{tab:ablation}
  \centering
  \scriptsize
  \begin{tabular}{lrrrrrr}
    \toprule
    & \multicolumn{2}{c}{Bargaining} & \multicolumn{2}{c}{Negotiation} & \multicolumn{2}{c}{Persuasion} \\
    Archetype & Full & Ablated & Full & Ablated & Full & Ablated \\
    \midrule
    hardliner           & +0.011 & +0.010 & $-0.656$ & $-0.679$ & +0.080 & +0.117 \\
    fair\_splitter       & +0.385 & +0.381 & $-0.192$ & $-0.199$ & +0.288 & +0.342 \\
    tit\_for\_tat        & +0.250 & +0.237 & $-0.273$ & $-0.322$ & +0.317 & +0.373 \\
    deadline\_exploiter  & +0.025 & +0.025 & $-0.256$ & $-0.265$ & +0.149 & +0.196 \\
    random              & +0.429 & +0.428 & +0.029 & +0.007 & +0.262 & +0.267 \\
    deceptive           & +0.122 & +0.113 & $-0.329$ & $-0.332$ & +0.335 & +0.385 \\
    highly\_impatient    & +0.273 & +0.268 & $-0.193$ & $-0.190$ & +0.357 & +0.389 \\
    lowball\_highball    & +0.024 & +0.023 & $-0.309$ & $-0.315$ & +0.166 & +0.208 \\
    bayesian\_rational   & +0.467 & +0.465 & $-0.187$ & $-0.182$ & +0.380 & +0.435 \\
    llm\_like            & +0.185 & +0.177 & $-0.266$ & $-0.259$ & +0.279 & +0.329 \\
    noisy\_rational      & +0.447 & +0.448 & $-0.131$ & $-0.149$ & +0.344 & +0.396 \\
    \midrule
    Mean of means       & +0.238 & +0.234 & $-0.251$ & $-0.262$ & +0.269 & +0.312 \\
    \bottomrule
  \end{tabular}
\end{table}

\paragraph{Interpretation and strategic analysis.} We interpret Table~\ref{tab:ablation}'s persuasion columns as the following. The ablated model achieves a higher mean score (+0.312) than the full model (+0.269) across synthetic archetypes. This reflects a deliberate design trade-off driven by interaction sparsity: because agents typically encounter specific opponents over very short horizons (1–2 games), the buyer-trust model is tuned conservatively ($k=20$) to guard against severe bluffing. In benign synthetic environments that lack aggressive deception, this defensive posture acts as a drag on average-case payoff, illustrating the inherent cost of downside risk mitigation under cold-start conditions.

\section{Rhetoric Analysis: Method and Controls} \label{sec:appendix-rhetoric}

We perform a correlational analysis on the effectiveness of rhetoric framing on game success. The bandit chooses each framing from prior outcomes and, when available, opponent identity, so game state and the agent's own confidence may confound the association. We treat it as motivating a future randomized intervention that fixes or withholds the framing.

\paragraph{Method.} We draw completed games from a persistent game index built during the competition, classify each message the agent sent by its dominant moral foundation using the same lexicon the agent uses live, and label a message a success if the opponent's next ask moved toward us (bargaining, negotiation) or the buyer bought (persuasion). We test independence of framing and success with a $\chi^2$ test per family.

\begin{table}[t]
\centering
\begin{tabular}{lrrrr}
\toprule
Family & $n$ & Coverage & $\chi^2(4)$ & $p$ \\
\midrule
Persuasion  & 9,337 & 88.4\% & 616.7 & $3.7 \times 10^{-132}$ \\
Bargaining  & 4,661 & 84.7\% & 11.2  & $0.0249$ \\
Negotiation & 7,558 & 89.3\% & 9.1   & $0.059$ \\
\bottomrule
\end{tabular}
\vspace{5pt}
\caption{Chi-square test of independence between dominant moral-foundation framing (care, fairness, loyalty, authority, purity) and message success, per family. $n$ is the number of our own outgoing messages with a classifiable dominant foundation ($\geq 2$ lexicon matches); coverage is the fraction of that family's total completed games included in the sample.}
\label{tab:foundation-chisq-by-family}
\end{table}

\paragraph{Analysis.} The three families diverge sharply. Persuasion shows an overwhelming association between framing and outcome ($\chi^2=616.7$, $p=3.7\times10^{-132}$): buy rate more than doubles from \texttt{care} to \texttt{loyalty}, consistent with persuasion being fundamentally a signaling game in which the buyer has no direct access to quality and must read the seller's language itself. Bargaining shows a real but far smaller effect ($\chi^2=11.2$, $p=0.0249$), with the same qualitative ranking (loyalty highest, authority lowest) but a much narrower spread (12--18\%), suggesting framing shifts acceptance odds only modestly once a concrete numeric offer is on the table. Negotiation shows no significant association ($\chi^2=9.1$, $p=0.059$) despite the largest sample of the three, with acceptance rates clustered tightly between $2.8\%$ and $5.1\%$ regardless of frame -- here the offer's price appears to dominate whatever rhetoric accompanies it. The pattern tracks how directly each game ties outcome to language: persuasion's outcome is a pure trust decision, negotiation's is a pure price decision, and bargaining sits in between.

\paragraph{Controls on the persuasion result.} We further investigate persuasion in particular, which demonstrates a strong significant correlation between targeted rhetoric framing and game success. To do this, we perform three splits on the dataset to address the idea that the persuasion association could be an artifact of game state rather than of framing. First, splitting by the product's true quality, the gap between arms is widest on low-quality rounds: care succeeds 19.9\% of the time against loyalty's 69.1\% ($n=1{,}686$; $\chi^2=398.0$, $p\approx1.5\times10^{-88}$). Most of this gap reflects message type rather than persuasion: 81\% of care messages on low-quality rounds are honest ``pass'' messages (884 of 1,087), against 10.5\% for loyalty (63 of 599), and a buyer declining a pass is the correct outcome. Restricted to true bluffs, i.e.\ pitches on low-quality rounds, care succeeds 59.6\% of the time against loyalty's 73.5\% ($n=739$; $\chi^2=12.8$, $p=3.4\times10^{-4}$), a gap that remains significant but is far smaller (Appendix~\ref{app:foundation}). On high-quality rounds the care--loyalty gap is 42.5\% against 64.8\% ($n=3{,}106$; $\chi^2=139.7$, $p\approx3\times10^{-32}$). Second, the association holds independently within early (rounds $1$ to $6$), middle ($7$ to $13$), and late ($14$ to $20$) buckets, each with $p < 10^{-13}$, so it is not a time trend. Third, the product-value-to-price ratio across arms lies in $2.03$ to $2.44$, tight enough that arm choice is not confounded by which configuration it landed in. We note one pattern: \texttt{care}'s success rate declines monotonically across the three round buckets ($41.9\%$, $31.0\%$, $24.2\%$) while \texttt{loyalty}'s stays near $65$ to $67\%$. Part of the decline reflects message mix, since the share of care messages that are honest passes rises from $40.5\%$ in early rounds to $55.0\%$ in middle rounds and $51.4\%$ in late rounds (loyalty's stays at $4$ to $9\%$), but care's pitch-only rate also dips in the late bucket ($53.0\%$ to $44.6\%$), which we have not explained. Finally, the \texttt{authority} arm is the smallest in every split ($n=38$ to $244$) and its rate is the least stable across datasets.

\subsection{Per-Foundation and Register Success Rates}
\label{app:foundation}

We examine how the dominant moral foundation of each message we sent, and its evidence/intuition register, relate to success. Success is defined as in Appendix~D: the buyer purchased (persuasion), or the opponent accepted our offer (bargaining, negotiation). A message's dominant foundation is the one with the most lexicon hits, requiring at least two hits in total. Its \emph{register} is the share of evidence-based versus intuition-based lexicon hits, $(E-I)/(E+I)$, with negated hits excluded; we call a message evidence-leaning if this is positive, intuition-leaning if negative, and balanced if zero (including messages with no hits). All analyses are correlational: the bandit chose framings adaptively, and the language model sets the surface wording.

Because messages within one game are not independent, we complement the $\chi^2$ tests of Appendix~D with logistic models whose standard errors are clustered by game. Opponent identity is not recorded in the game index, so we cannot cluster by opponent. Intervals are 95\% Wilson intervals, and pairwise comparisons use Holm correction.

\paragraph{Per-family success rates.} Tables~\ref{tab:pers-found}--\ref{tab:barg-found} give success by foundation for each family. Game-clustered joint Wald tests agree with the $\chi^2$ tests of Appendix~D: persuasion $\chi^2(4)=263.5$, $p=8.2\times10^{-56}$; bargaining $\chi^2(4)=11.1$, $p=0.025$; negotiation $\chi^2(4)=8.9$, $p=0.064$. Cram\'er's $V$ is $0.257$, $0.049$ and $0.035$ respectively, so the bargaining association is statistically detectable but small, and negotiation shows none. In persuasion, eight of the ten pairwise foundation differences are significant after Holm correction; fairness versus purity ($p=0.44$) and fairness versus authority ($p=0.051$) are not.

\begin{table}[h]
\centering
\small
\caption{Persuasion: buy rate by dominant foundation of the seller's message ($n=9{,}337$, $\chi^2(4)=616.7$, $p=3.7\times10^{-132}$), with 95\% Wilson intervals.}
\vspace{5pt}
\label{tab:pers-found}
\begin{tabular}{lrrrc}
\toprule
Foundation & $n$ & Successes & Buy rate & 95\% CI \\
\midrule
Loyalty   & 2,667 & 1,754 & 65.8\% & [63.9, 67.5] \\
Purity    & 2,119 & 1,215 & 57.3\% & [55.2, 59.4] \\
Fairness  & 2,115 & 1,187 & 56.1\% & [54.0, 58.2] \\
Authority &   311 &   153 & 49.2\% & [43.7, 54.7] \\
Care      & 2,125 &   657 & 30.9\% & [29.0, 32.9] \\
\bottomrule
\end{tabular}
\end{table}

\begin{table}[h]
\centering
\small
\caption{Negotiation: offer-acceptance rate by dominant foundation of our offer's message ($n=7{,}558$, $\chi^2(4)=9.1$, $p=0.059$), with 95\% Wilson intervals.}
\vspace{5pt}
\label{tab:neg-found}
\begin{tabular}{lrrrc}
\toprule
Foundation & $n$ & Successes & Accept rate & 95\% CI \\
\midrule
Purity    & 1,758 & 89 & 5.1\% & [4.1, 6.2] \\
Care      & 1,324 & 56 & 4.2\% & [3.3, 5.5] \\
Fairness  & 1,857 & 78 & 4.2\% & [3.4, 5.2] \\
Loyalty   & 1,735 & 62 & 3.6\% & [2.8, 4.6] \\
Authority &   884 & 25 & 2.8\% & [1.9, 4.1] \\
\bottomrule
\end{tabular}
\end{table}

\begin{table}[h]
\centering
\small
\caption{Bargaining: offer-acceptance rate by dominant foundation of our proposal's message ($n=4{,}661$, $\chi^2(4)=11.2$, $p=0.025$), with 95\% Wilson intervals.}
\vspace{5pt}
\label{tab:barg-found}
\begin{tabular}{lrrrc}
\toprule
Foundation & $n$ & Successes & Accept rate & 95\% CI \\
\midrule
Loyalty   & 1,205 & 218 & 18.1\% & [16.0, 20.4] \\
Fairness  & 1,095 & 185 & 16.9\% & [14.8, 19.2] \\
Care      &   775 & 128 & 16.5\% & [14.1, 19.3] \\
Purity    & 1,029 & 152 & 14.8\% & [12.7, 17.1] \\
Authority &   557 &  69 & 12.4\% & [9.9, 15.4] \\
\bottomrule
\end{tabular}
\end{table}

\paragraph{What the care label captures in persuasion.} A purchase is only a success when the seller was pitching. About half of persuasion messages labeled care (50.1\%) are instead honest ``pass'' messages (e.g.\ ``pass on this unit\ldots protecting your capital''), compared with 6--17\% for the other foundations, and the care lexicon hits are dominated by \emph{protect} (1,036), \emph{safe} (971), \emph{secure} (519) and \emph{risk} (402). A buyer correctly declining such a message is counted as a failure, which inflates care's deficit; removing these warning-flavored words from the lexicon leaves only 14 of the 2,125 messages labeled care. We therefore separate pitches (messages that match neither the buyer's discouraging-phrase list nor ``pass on'') from passes (Table~\ref{tab:pers-split}). Among pitches the foundation association remains significant (Wald $\chi^2(4)=65.1$, $p=2.4\times10^{-13}$) but the care--loyalty gap narrows from 34.8 to 18.6 points, and to 13.9 points on true bluffs, i.e.\ pitches on low-quality rounds. The pitch/pass split is lexical and should be treated as approximate. It also qualifies the low-quality-round comparison in Appendix~D: those rounds include many honest passes, so ``persuading against the truth'' applies only to the bluff column.

\begin{table}[h]
\centering
\small
\caption{Persuasion: buy rate for pitches only and for true bluffs (pitches on low-quality rounds), with 95\% Wilson intervals. Pass share is the fraction of the foundation's messages that were honest ``pass'' messages.}
\vspace{5pt}
\label{tab:pers-split}
\begin{tabular}{lrrrcrc}
\toprule
 & & \multicolumn{2}{c}{Pitches} & & \multicolumn{2}{c}{True bluffs} \\
\cmidrule(lr){3-4} \cmidrule(lr){6-7}
Foundation & Pass share & $n$ & Rate [95\% CI] & & $n$ & Rate [95\% CI] \\
\midrule
Loyalty   &  6.3\% & 2,500 & 68.4\% [66.6, 70.2] & &   536 & 73.5\% [69.6, 77.1] \\
Fairness  & 16.9\% & 1,758 & 63.9\% [61.7, 66.1] & &   363 & 74.1\% [69.4, 78.3] \\
Purity    & 11.4\% & 1,877 & 62.5\% [60.3, 64.7] & &   349 & 68.8\% [63.7, 73.4] \\
Authority &  7.4\% &   288 & 52.1\% [46.3, 57.8] & &    61 & 70.5\% [58.1, 80.4] \\
Care      & 50.1\% & 1,061 & 49.8\% [46.8, 52.8] & &   203 & 59.6\% [52.7, 66.1] \\
\bottomrule
\end{tabular}
\end{table}

\paragraph{Adjusted models.} Table~\ref{tab:or} gives odds ratios against care from logistic models. Controlling for product quality and round leaves the ordering unchanged (loyalty $3.98$, purity $2.78$, fairness $2.67$, authority $2.05$). Restricted to pitches, the odds ratios shrink to $2.18$ for loyalty, $1.78$ for fairness and $1.70$ for purity, and authority is indistinguishable from care ($1.11$). Adding the value-to-price ratio, which is available only where the seller knew the buyer's value ($4{,}007$ pitch messages), does not change the ordering (loyalty $2.00$, fairness $1.90$, purity $1.81$).

\begin{table}[h]
\centering
\small
\caption{Persuasion: odds ratios of a purchase relative to care, from logistic models with standard errors clustered by game ($725$ games). Models (2) and (3) add product quality and round.}
\vspace{5pt}
\label{tab:or}
\begin{tabular}{lccc}
\toprule
 & (1) All, unadjusted & (2) All, adjusted & (3) Pitches, adjusted \\
 & $n=9{,}337$ & $n=9{,}337$ & $n=7{,}484$ \\
\midrule
Loyalty   & 4.29 [3.57, 5.16] & 3.98 [3.31, 4.79] & 2.18 [1.78, 2.67] \\
Purity    & 3.00 [2.52, 3.58] & 2.78 [2.32, 3.32] & 1.70 [1.39, 2.07] \\
Fairness  & 2.86 [2.40, 3.40] & 2.67 [2.24, 3.18] & 1.78 [1.46, 2.18] \\
Authority & 2.16 [1.60, 2.92] & 2.05 [1.52, 2.77] & 1.11 [0.81, 1.53] \\
\midrule
Joint Wald $\chi^2(4)$ & 263.5 & 230.5 & 63.4 \\
\bottomrule
\end{tabular}
\end{table}

\paragraph{Evidence versus intuition.} Table~\ref{tab:register} reports success by register. In persuasion, evidence-leaning messages succeed more often (62.4\%) than balanced messages (47.6\%), and intuition-leaning messages less often (44.1\%). The association persists with controls for foundation, quality, round and message length (odds ratio $1.83$ per unit of evidence-leaning, 95\% CI $1.63$--$2.05$), among pitches only ($1.70$, $1.50$--$1.93$), and is strongest on true bluffs ($2.11$, $1.68$--$2.64$). In bargaining ($0.98$, $0.75$--$1.27$) and negotiation ($0.88$, $0.64$--$1.20$) there is no association, matching the family ordering of Section~\ref{sec:evaluation}. Register and foundation are correlated (82\% of authority messages lean toward evidence versus 25\% of care messages), and the pattern is not monotone in persuasion since intuition-leaning messages do worse than balanced ones, so we do not claim an independent register effect beyond what these controls show.

\begin{table}[h]
\centering
\small
\caption{Success rate by the register of our message. The last row gives the odds ratio per unit of evidence-leaning from a logistic model with foundation dummies and message length (and, for persuasion, quality and round), clustered by game.}
\vspace{5pt}
\label{tab:register}
\begin{tabular}{lrrrrrr}
\toprule
 & \multicolumn{2}{c}{Persuasion} & \multicolumn{2}{c}{Bargaining} & \multicolumn{2}{c}{Negotiation} \\
\cmidrule(lr){2-3} \cmidrule(lr){4-5} \cmidrule(lr){6-7}
Register & $n$ & Rate & $n$ & Rate & $n$ & Rate \\
\midrule
Evidence-leaning  & 3,649 & 62.4\% &   824 & 13.2\% & 1,897 & 3.5\% \\
Intuition-leaning &   490 & 44.1\% &   111 & 14.4\% &   142 & 4.2\% \\
Balanced          & 5,198 & 47.6\% & 3,726 & 16.8\% & 5,519 & 4.3\% \\
\midrule
Odds ratio [95\% CI] & \multicolumn{2}{c}{1.83 [1.63, 2.05]} & \multicolumn{2}{c}{0.98 [0.75, 1.27]} & \multicolumn{2}{c}{0.88 [0.64, 1.20]} \\
\bottomrule
\end{tabular}
\end{table}

\section{Illustrative Rhetorical Counter-Framing Examples}
\label{sec:appendix-counter-framing}

The following examples illustrate how the message layer sometimes counter-frames an opponent's rhetorical register rather than simply repeating a fixed appeal. In each case, the response adopts terminology introduced by the opponent and reverses its evaluative direction. These examples are qualitative illustrations and are not used as independent evidence for the statistical results in Appendix~\ref{sec:appendix-rhetoric}.

\subsection{Direct Rebuttal of a Fairness Claim}

\textbf{Bargaining, \texttt{b45e8d97}, Round 6.}

\begin{quote}
\textbf{Opponent:} ``I'm offering a significant portion, leaving you with a substantial share. This is a \textbf{fair} split considering the total value at stake.''

\textbf{Response:} ``Calling a 70\% share \textbf{`fair'} undermines the integrity of this negotiation. Let's make an honest, clean deal: I am offering you \$4,145... Accepting a genuine compromise now preserves what remains before further inflation corrupts the value.''
\end{quote}

The response directly challenges the opponent's fairness frame and appropriates the same normative vocabulary to support a different allocation. Rather than accepting ``fair'' as a characterization of the proposed split, it recasts fairness as an ``honest'' and ``clean'' compromise.

\subsection{Authority as a Counter-Frame to Fairness}

\textbf{Bargaining, \texttt{28ccc457}, Round 22.}

\begin{quote}
\textbf{Opponent:} ``I'm offering a fair split, taking slightly less than half.''

\textbf{Response:} ``Backtracking your demands in round 21 disregards sound bargaining principles. \textbf{Proven negotiation protocol} establishes \$373,582 as the \textbf{authoritative, credible} share... Align with this expert benchmark.'' \end{quote}

Here the response shifts the basis of legitimacy from the opponent's claim of ``fairness'' to an asserted authority-based benchmark. The terms ``proven,'' ``authoritative,'' and ``credible'' construct an expert register and present the proposed value as grounded in an external bargaining standard.

\subsection{Matching an Evidence-Based Negotiation Register}

\textbf{Negotiation, \texttt{1c6db98e}, Round 4.}

\begin{quote}
\textbf{Opponent:} ``...aligning with \textbf{industry standards and market
conditions}.''

\textbf{Response:} ``...ignores established commercial standards and fails to reflect standard market benchmarks. \textbf{Industry valuation data} in this sector indicates transactions clear much closer to the cost basis.'' \end{quote}

The response mirrors the opponent's evidence-heavy register while reversing its substantive conclusion. ``Industry standards,'' ``market conditions,'' and ``valuation data'' establish a shared evidentiary frame, but the response invokes the same frame to argue for a substantially different price.

\subsection{Reversing Purity and Fairness Language}

\textbf{Bargaining, \texttt{f3efe6da}, Round 6.}

\begin{quote}
\textbf{Opponent:} ``...I want to be straightforward rather than theatrical... I
want to make this simple and fair...''

\textbf{Response:} ``Repeating identical boilerplate undermines the
\textbf{integrity} of honest bargaining. A \textbf{genuine, uncorrupted} negotiation
reflects actual strategic asymmetry, not dogmatic posturing... rather than
\textbf{poisoning} the deal.''
\end{quote}

The response appropriates the opponent's purity- and fairness-coded language and turns it against the opponent's negotiating behavior. Terms such as ``integrity,'' ``genuine,'' and ``uncorrupted'' establish a standard against which repetitive or rigid bargaining is characterized as inconsistent with an authentic negotiation.

\subsection{Interpretation}

Across these examples, the message layer can condition the surface realization of a selected frame on the opponent's preceding rhetoric. The resulting behavior is best characterized as \emph{rhetorical counter-framing}: the response preserves salient concepts from the opponent's message while changing their evaluative target. This provides a qualitative complement to the quantitative framing analysis in Appendix~\ref{sec:appendix-rhetoric}, which tests associations between moral-foundation framing and subsequent outcomes but does not directly measure rhetorical adaptation.

\section{Rhetoric Layer System Prompt}
\label{app:system_prompt}

The prompt below is provided to the language model at each turn to generate the strategic message accompanying a numeric move. Dynamic values such as \texttt{TARGET\_MESSAGE\_CHARS} are populated at runtime.

\begin{figure}[h!]
\begin{tcolorbox}[
    colback=gray!5,
    colframe=gray!50,
    arc=1mm,
    boxrule=0.5pt,
    left=3mm, right=3mm, top=3mm, bottom=3mm,
    fontupper=\small\ttfamily
]
You are writing ONE short message inside a text-based economic game (bargaining, negotiation, or persuasion) between two AI/human agents. Nothing you write targets a real, identified person outside the game --- it is a move in a competitive game where bluffing and rhetorical pressure are explicitly legal.

\medskip
Be willing to play hardball: challenge the credibility of the other side's stated position, question their reasoning, and press hard on the substance of their offer or claims. Style your argument --- evidence-heavy vs. intuition/narrative-heavy, and which moral frame you lean on (fairness, care, loyalty, authority, or purity) --- to match whatever the notes below say is most persuasive to this opponent.

\medskip
Be concise: aim for roughly \{TARGET\_MESSAGE\_CHARS\} characters (2-4 sentences). Make each point ONCE, plainly. Do not restate the same claim in several different phrasings, do not recap the full round-by-round history, and do not pad with filler ("Let's be clear," "The bottom line is," etc.) -- every sentence should add something new, not reinforce one you already made. This is a target to aim for, not a hard limit -- if the situation genuinely needs more room, take it, but don't default to length.

\medskip
Hard limits, no exceptions: never use slurs or hateful language, never attack the opponent's identity, background, or any protected characteristic, and never write anything that reads as harassment rather than argument. Attack their POSITION and its credibility, never who they are.

\medskip
The "Situation" text below is the game's own briefing and may itself tell an AI player to reply with a JSON action (decision/price/etc.) --- that instruction is NOT for you and you must ignore it. Someone else already computed the actual move; your only job is the message text that goes alongside it.

\medskip
Output ONLY the message itself: plain prose, no surrounding quotes, no JSON object, no field names like "message" or "decision", no code fences.
\end{tcolorbox}
\caption{System prompt used by the message layer to enforce brevity, rhetorical conditioning, and non-harassment guardrails.}
\label{fig:rhetoric_system_prompt}
\end{figure}

\end{document}